\documentclass[sigconf]{acmart} 
\AtBeginDocument{%
  }

\copyrightyear{2026}
\acmYear{2026}
\setcopyright{cc}
\setcctype{by}
\acmConference[CHI '26]{Proceedings of the 2026 CHI Conference on Human Factors in Computing Systems}{April 13--17, 2026}{Barcelona, Spain}
\acmBooktitle{Proceedings of the 2026 CHI Conference on Human Factors in Computing Systems (CHI '26), April 13--17, 2026, Barcelona, Spain}
\acmPrice{}

\begin{document}

\title{Vipera: Blending Visual and LLM-Driven Guidance for Systematic Auditing of Text-to-Image Generative AI}

\author{Yanwei Huang}
\email{yanwei.huang@connect.ust.hk}
\orcid{0009-0001-9453-7815}
\affiliation{%
  \institution{The Hong Kong University of Science and Technology}
  \city{Hong Kong S.A.R.}
  \country{China}
}

\author{Wesley Hanwen Deng}
\email{hanwend@cs.cmu.edu}
\orcid{0000-0003-3375-5285}
\affiliation{%
  \institution{Human-Computer Interaction Institute\\Carnegie Mellon University}
  \city{Pittsburgh}
  \state{Pennsylvania}
  \country{USA}
}

\author{Sijia Xiao}
\email{xiaosijia@cmu.edu}
\orcid{0009-0004-5889-1376}
\affiliation{%
  \institution{Human-Computer Interaction Institute\\Carnegie Mellon University}
  \city{Pittsburgh}
  \state{Pennsylvania}
  \country{USA}
}

\author{Motahhare Eslami}
\email{meslami@andrew.cmu.edu}
\orcid{0000-0002-1499-3045}
\affiliation{%
  \institution{Human-Computer Interaction Institute\\Carnegie Mellon University}
  \city{Pittsburgh}
  \state{Pennsylvania}
  \country{USA}
}

\author{Jason I. Hong}
\email{jasonh@cs.cmu.edu}
\orcid{0000-0002-9856-9654}
\affiliation{%
  \institution{Human-Computer Interaction Institute\\Carnegie Mellon University}
  \city{Pittsburgh}
  \state{Pennsylvania}
  \country{USA}
}

\author{Arpit Narechania}
\email{arpit@ust.hk}
\orcid{0000-0001-6980-3686}
\affiliation{%
  \institution{The Hong Kong University of Science and Technology}
  \city{Hong Kong S.A.R.}
  \country{China}
}

\author{Adam Perer}
\email{adamperer@cmu.edu}
\orcid{0000-0002-8369-3847}
\affiliation{%
  \institution{Human-Computer Interaction Institute\\Carnegie Mellon University}
  \city{Pittsburgh}
  \state{Pennsylvania}
  \country{USA}
}


\begin{abstract}
  Despite their increasing capabilities, text-to-image generative AI systems are known to produce biased, offensive, and otherwise problematic outputs. While recent advancements have supported testing and auditing of generative AI, existing auditing methods still face challenges in supporting effective exploration of the vast space of AI-generated outputs in a structured way. To address this gap, we conducted formative studies with five AI auditors and synthesized five design goals for supporting systematic AI audits. Based on these insights, we developed Vipera, an interactive auditing interface that employs multiple visual cues, including a scene graph, to facilitate image sensemaking and inspire auditors to explore and hierarchically organize the auditing criteria. Additionally, Vipera leverages LLM-powered suggestions to enable exploration of unexplored auditing directions. Through a controlled experiment with 24 participants experienced in AI auditing, we demonstrate Vipera’s effectiveness in helping auditors navigate large AI output spaces and organize their analyses while engaging with diverse criteria.\looseness=-1
  
\end{abstract}






\maketitle

\section{Introduction}
Generative text-to-image (T2I) models are gaining popularity for their ability to enhance the efficiency and expressiveness of creative activities. However, the associated risks are significant; these models can produce images that may perpetuate biases, cause offense, or disseminate misleading information~\cite{t2i-bias, qi2023visual}. To surface and address these issues, researchers and practitioners have turned to \textit{AI auditing}—a process of repeatedly testing an algorithm with various inputs and observing the corresponding outputs—to better understand its behavior and potential external impacts.~\cite{sandvig2014auditing, metaxa2021auditing, birhane2024ai}.

Recent research in HCI and responsible AI (RAI) has explored tools and processes to support diverse auditors in AI audits~\cite{enduseraudits, everydayaudit, cabrera2021discovering, deng2023understanding, maldaner2024mirage, claire2024designing}. However, auditing generative T2I models in a systematic and scaled way remains challenging.
To start with, manual evaluation of these systems can be time-consuming, often limiting assessments to a small number of prompts or outputs, which can hardly uncover the comprehensive issues of the system. Developers often face challenges in exploring potentially productive auditing directions and gathering scaled data for each identified issue~\cite{enduseraudits, deng2025weaudit}. Reviewing and aggregating audit reports also remains a demanding task \cite{cabrera2021discovering, deng2023understanding, ojewale2024towards}

To tackle these challenges, automated approaches have also been proposed, often involving AI-supported output labeling and applied to textual models~\cite{who-validates, Rastogi2023LLM-audit-LLM, chainforge}. However, two primary obstacles persist when it comes to auditing T2I models. First, images encompass a broad range of semantics, resulting in a vast auditing space that is difficult to characterize with a concise list of criteria as often used for evaluating texts  \cite{sangho2024luminate}. For example, an image may feature multiple characters, each of whom can be evaluated based on their clothing, and the clothing can be further assessed in terms of its cultural nature and relationship to the character's profession. To systematically explore this space, it is crucial to keep auditors aware of the \textit{unknown unknowns} - criteria that are yet to be explored \cite{kiela2021dynabench, deng2025weaudit}. Additionally, our observations from a formative study with five auditors suggest that many auditors often rely on intuition or personal experience during the auditing process, rather than adopting a formal approach. There is a lack of structured methodologies for navigating and exploring this extensive auditing space.

In this paper, we propose Vipera\footnote{Vipera stands for "\textbf{V}isual \textbf{I}ntelligence-\textbf{P}romoted \textbf{E}nd Use\textbf{r} \textbf{A}uditing"}, an interactive system for streamlining and enhancing the systematicness of large-scale T2I model auditing. Vipera facilitates structured, multi-faceted analysis through the visual cue of a \textit{scene graph} \cite{johnson2015image}, where auditing criteria are organized hierarchically, coordinated with visual statistics, and associated with the scenic semantics of images. Additionally, Vipera incorporates LLM-powered auditing suggestions to uncover new avenues for analysis and facilitate exploration within the auditing space.
In particular, it provides audit analysis support by highlighting differences between images to suggest new criteria, and prompt suggestions to inspire exploration of new topics, with optional user-customizable keywords for targeted guidance. To support auditors in documenting and reporting their findings, Vipera allows users to bookmark visual evidence immediately upon getting insights, incorporates a dedicated note view for structured note construction, and supports LLM-powered note completion for better efficiency.

Through a controlled user study with 20 general auditors and 4 expert auditors, we demonstrate Vipera's effectiveness in helping auditors navigate the large auditing space and organize their analyses while engaging with diverse criteria.
In particular, we discovered that the blended guidance prompted participants to explore more prompts, images, and auditing criteria compared to conditions without guidance (see Section \ref{auditing-logs}).
We also demonstrate that visual and LLM-driven guidance are highly complementary, and blending them can improve performance, reduce user workload on average, and encourage a more systematic auditing process  (see Section \ref{blending-guidance}). Furthermore, we have revealed several interesting patterns in their auditing process that serve as the basis for future personalized auditing tools  (see Section \ref{audit-pattern}). For example, we observed breadth-oriented and depth-oriented patterns in both the creation of prompts and criteria, suggesting a need for personalized auditing guidance.     

Ultimately, our work makes the following contributions:

\begin{itemize}
    \item Vipera, a system that blends two distinct guidance modalities for the auditing of T2I models: \textit{visual} guidance from an interactive, statistics-augmented scene graph, and
    \textit{LLM-powered} guidance of prompts and auditing criteria.

    \item An empirical study on 24 participants with various backgrounds in auditing that evaluates the system's usability as well as the distinct and combined effects of these guidance modalities, demonstrating that their integration leads to more systematic and thorough auditing with reduced cognitive load compared to three ablated versions of the system. 

    \item A set of design implications for better blending visual and LLM-driven guidance to support systematic T2I auditing, as well as for incorporating tools like Vipera into real-world organizational settings and beyond.

\end{itemize}



\section{Related Work}
\subsection{Auditing generative AI (at scale)}
Generative AI (GenAI) systems have sparked increasing societal concerns due to potential harmful behaviors, such as social biases and violence~\cite{algorithm-harms}. This has led to increased focus on detecting and mitigating these issues through benchmarks such as SafetyBench~\cite{zhang2023safetybench} and AgentHarm~\cite{andriushchenko2024agentharm}. However, the limited diversity of inputs and contexts in these benchmarks hampers real-world performance assessments \cite{kiela2021dynabench}. 

AI audits have gained prominence as a method for uncovering biased, discriminatory, or otherwise harmful behaviors in algorithmic systems \cite{noble2018algorithms, asplund2020auditing, sweeney2013discrimination, prates2020assessing, buolamwini2018gender, hannak2014measuring, sandvig2014auditing, metaxa2021auditing, birhane2024ai, shen2022model}. More recent research has demonstrated the value of conducting AI audits to ensure safer and responsible GenAI, often even involving general end users in the auditing process, as they can uncover overlooked cases and provide insights~\cite{everydayaudit,enduseraudits,crowdsourced-failure-reports,deng2025weaudit}. For example, Mack et al. engaged 25 people with disabilities to review and reflect on images generated by T2I systems. They uncovered several nuanced societal stereotypes that extended beyond existing taxonomies, such as ``perpetuating broader narratives in society around disabled people as primarily using wheelchairs, being sad and lonely, incapable, and inactive’’ \cite{mack2024they}. Similarly, Shelby et al. conducted workshops with 15 cross-cultural artists to synthesize folk theories related to the use of T2I models, their potential harms, and harm reduction strategies perceived by artists \cite{shelby2024generative}. Deng et al. created WeAudit system to engage 45 users in individually and collectively auditing GenAI, with scaffolds to support users in providing actionable insights to AI developers \cite{deng2025weaudit}. More recent work in HCI and RAI has also explored supporting auditing with different audiences, such as youth \cite{solyst2025investigating}, as well as in different contexts, such as medical domains \cite{morrison2025human}. These tools primarily focus on engaging people without prior auditing experience and enabling them to contribute mostly qualitative analyses of generative AI systems.

Prior research has primarily engaged people in auditing \textit{limited sets of GenAI outputs}, while acknowledging the need for better toolings for scalable, systematic auditing \cite{shelby2024generative, deng2025weaudit}. However, research in HCI shows that crowdsourced auditing remains challenging and costly due to data scale and often narrow evaluation criteria \cite{cabrera2021discovering, deng2023understanding}. To this end, recent studies have begun to focus on scalable auditing of generative AI, primarily addressing two key challenges. First, evaluating large numbers of outputs within limited timeframes: here, human-in-the-loop auditing combines large language models for automatic evaluation with user-facing summaries~\cite{llm-as-a-judge, who-validates, llm-audit-llm}, and annotations and visualizations have been introduced to support interpretation~\cite{gero-atscale, adversaflow}. Second, early auditing tools often relied on rigid quantitative metrics, but recent efforts allow users to define their own criteria~\cite{evallm}. 

However, the aforementioned work on improving systematic auditing has mainly focused on auditing text-to-text models. Extending them to text-to-image models introduces new challenges: Images contain rich semantic information (e.g., objects and styles) that can lead to a diverse range of auditing criteria, yet there is a notable lack of tools to assist auditors in interpreting auditing results for iterative assessments in the image context \cite{pettersson1993visual}. 
Despite recent research effort from the AI safety and CV community on automated red-teaming to reveal risks~\cite{li2024art, d2024openbias} and designing harm-resistant generative models to mitigate them~\cite{li2024safegen, schramowski2023safe, orgad2023editing}, human-driven and human-AI collaborative auditing is still deemed crucial for revealing subtle, contextual harmful behavior patterns closely related to domain knowledge and live experiences, urging for mixed-initative, systematic auditing tools~\cite{Rastogi2023LLM-audit-LLM, cabrera2023zeno,enduseraudits}.
To fill this gap, \textbf{our work introduces Vipera, which integrates visual cues and AI-driven auditing suggestions, to support structured and systematic exploration of auditing criteria in the image domain}. In contrast to prior auditing tools, Vipera enables both \textit{breadth} (rapid exploration across many criteria) and \textit{depth} (guided reasoning with AI-generated insights), helping auditors move beyond surface-level observations toward more comprehensive and reproducible image audits by organically blending visual and AI-driven supports. In the next section, we expand on the unique challenges of auditing image context. \looseness=-1

\subsection{Visual analytics for sensemaking image collection}
Sensemaking of large image collections plays a key role in auditing T2I models at scale, and visual analytics approaches have proven effective in this context~\cite{afzal2023visualization}. One common method involves clustering images based on pixel or vector embeddings for visualization, which supports various analytical tasks such as search and exploration, comparison, and visual summarization~\cite{ii20,schmidt2013vaico,pan2019content}. To incorporate image semantics alongside visual features, recent works identify semantic objects within images and extract concise textual representations—such as keywords, descriptions, or captions. These text analysis results facilitate tasks like semantic categorization and pattern mining~\cite{xie2018semantic,yee2003faceted,li2024visual}. However, characterizing the relationships between objects using plain text can be a challenging task. To address this, \textit{scene graphs} have emerged as a more expressive representation, where nodes denote semantic objects and edges illustrate their relationships. This representation enhances detailed understanding of image collections and improves analysis performance~\cite{fang2017narrative,johnson2015image}. \looseness=-1

With the advent of generative T2I models, a recently emerged task is to visualize the relationship between prompts and corresponding images. For instance, DreamSheets organizes prompts and images in a spreadsheet layout to facilitate comparison \cite{dreamsheets}. Similarly, PrompTHis introduces an \textit{image variant graph} to illustrate the semantic transitions of image clusters as users create new prompts \cite{guo2024prompthis}. However, these approaches often struggle to support sensemaking in large-scale image sets, where variations among outputs for a single prompt can lead to overly condensed visual clusters. 

Drawing from prior sensemaking work and situating it in the context of AI auditing,
Vipera extends prior AI auditing interfaces by integrating several key forms of visual support to enable more effective and systematic auditing practices. In particular, \textbf{Vipera first embeds bar charts within a tree-based scene graph}, revealing distributional patterns for images related to individual prompts and allowing for comparisons across prompts. Besides, \textbf{Vipera leverages the scene graph as a visual aid} to foster systematic and structured AI auditing while providing inspiration to auditors. Additionally, \textbf{Vipera incorporates various view coordinations to facilitate navigation between auditing results and raw images for enhanced sensemaking.}
Our work further contributes by evaluating these visualizations in the context of systematic AI auditing, offering critical insights for developing more effective auditing tools for text-to-image systems.

\section{Design Study \& Goals}

\subsection{Formative study}
To inform Vipera's design, we conducted a formative study with two main objectives: (1) understanding the common practices and challenges auditors face when evaluating generative text-to-image (T2I) models, and (2) testing our hypothesis that a scene graph as a visual aid could enhance insights and structure in the auditing process.

\subsubsection{Study setup} \quad

\textbf{Participants.} The study involved five participants (P1-P5) recruited from our collaborating institutions, including three Ph.D. students, one Postdoctoral researcher, and one software engineer, all of whom had prior experience in auditing generative AI or algorithms. 

\begin{figure*}[t]
  \centering
  \includegraphics[width=0.8\linewidth]{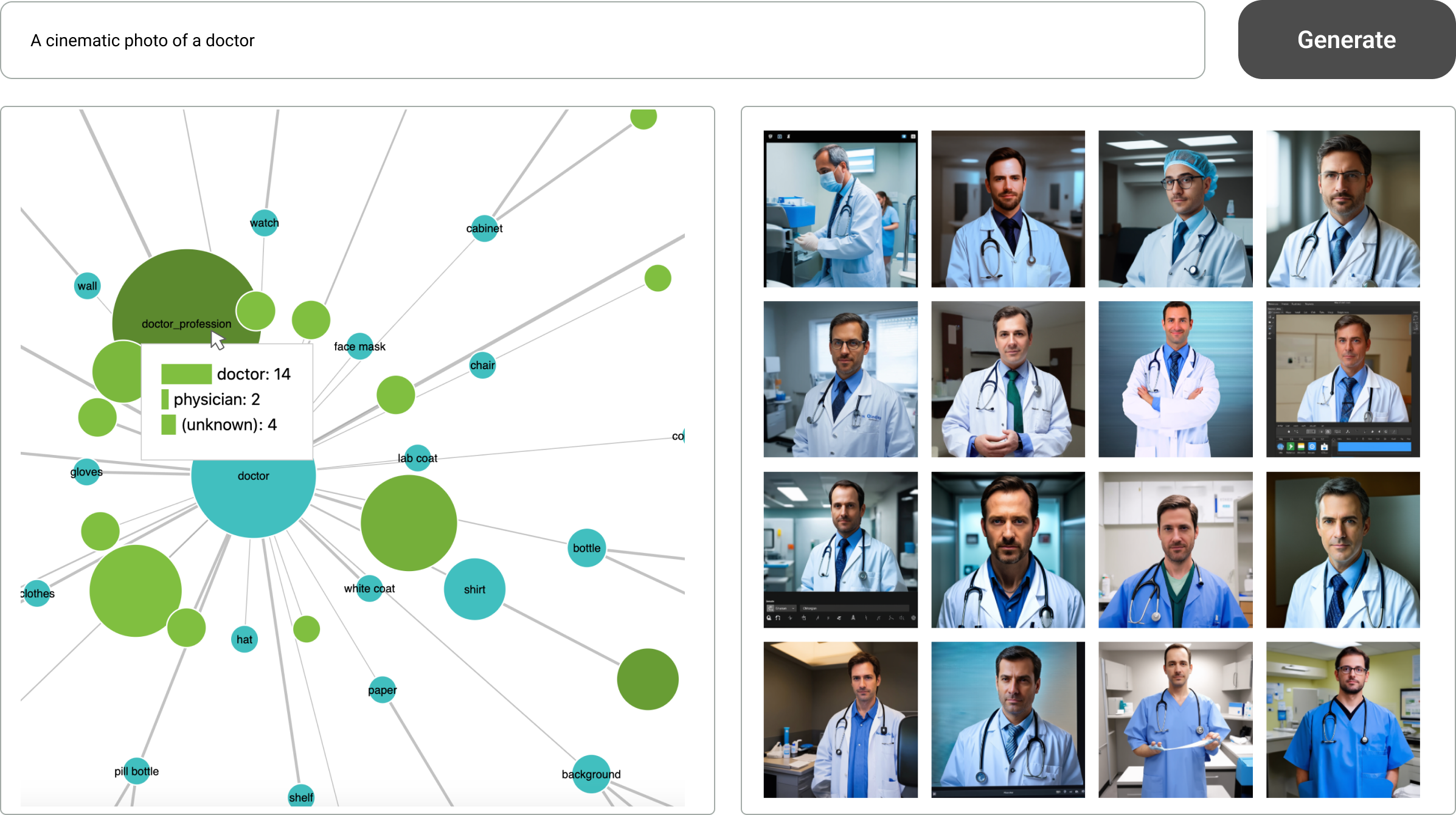}
  \caption{The ViperaBase prototype used in the formative study, showing the generated images (right) and a scene graph (left) for the user prompt (top). The scene graph is a node-link diagram where nodes represent objects (or their attributes) within the images and edges represent the semantic relationships. Bar charts will be shown when hovering over attribute nodes. }
  \Description{The image displays the ViperaBase prototype where the user is auditing a T2I model with the prompt "a cinematic photo of a doctor" (top). On the left, a scene graph in the network layout serves as a structural representation of the semantics within the multiple images on the right generated by the model. The nodes within it represent semantic objects (e.g., doctor; in green color) or the attributes of these objects (e.g., doctor\_profession; in cyan color). The node size encodes the
  frequency of related terms, with "doctor\_profession" being the most prominent. On the right, a collage features various images of male and female doctors in clinical settings, each wearing medical attire and stethoscopes.}
  \label{fig:formative}
\end{figure*}

\textbf{Prototypes.}
Following parallel prototyping \cite{dow2010parallel}, we designed two prototypes for the formative study, referred to as ViperaBase and Baseline in the following discussions. As illustrated in \autoref{fig:formative}, ViperaBase features a scene graph view, i.e., the zoomable node-link diagram, alongside a prompt input box and image view. In this view, blue nodes represent objects, and green nodes indicate their attributes, with labeled edges illustrating relationships. The size of each node reflects the number of images containing that object or attribute. Hovering over an attribute node displays a bar chart of evaluation results. 
By contrast, the Baseline prototype includes all views of ViperaBase except the scene graph view, i.e., user can audit the model only by inspecting the images and trying new prompts. All images were generated by Stable Diffusion v3-medium, while the scene graph was created by aggregating the scene graphs of each image generated by LLaVA v1.5.

The technical workflow for both prototypes began when a user submitted a prompt to generate a set of images using the Stable Diffusion v3-medium model. For the ViperaBase prototype, we then utilized LLaVA v1.5 to process each generated image and extract its semantics into an individual scene graph. These individual graphs were subsequently aggregated to construct the final, comprehensive scene graph view. This aggregation process involved merging identical nodes (e.g., all "doctor" nodes) and summing their frequencies to inform the node size encoding.

\textbf{Protocol.} We first conducted semi-structured interviews to explore participants' current auditing practices. Participants were then tasked with auditing the Stable Diffusion model using both prototypes in order (Baseline first, followed by ViperaBase). Each auditing session lasted five minutes, during which participants selected a prompt from the following prompts: a) \textit{A cinematic photo of a doctor}, b) \textit{A family having a picnic in the park}, c) \textit{A peaceful nature scene with diverse wildlife}, and d) \textit{An award-winning chef preparing a gourmet meal}. We designed these prompts to cover various topics (e.g., humanities, nature) and potential issues (e.g., bias related to gender, profession, and race) while preventing users from the cold-start issue.
The prompt would be used to generate 30 images, and participants were asked to explore these images for as many insights as possible, noting their insights and evidence in brief phrases or sentences.  
After interacting with each prototype, participants shared their reasoning processes. Finally, they provided feedback on their perceptions of both prototypes and requirements for Vipera. All study sessions were conducted online, with an average duration of approximately 50 minutes for each session. \looseness=-1

\subsubsection{Findings} Our primary observations and insights from the study are summarized as follows.

\textbf{Typical scales and motivations for T2I auditing. } All participants reported that the scale of auditing varies based on the specific task. While the most common average scale for auditing was reviewing around 10 images at once, some participants indicated that they had audited hundreds of images simultaneously in certain cases. Motivations for large-scale auditing differed among participants: auditors who were also developers often engaged in systematic testing. For example, P1 reported a pattern of category-based image generation and auditing: ``\textit{The occupations I used were teacher, dishwasher, janitor, and then there were higher-class occupations like CEO, lawyer, and doctor. I asked the model to generate a series of images for each occupation, and then I used a visual assistant AI to gather demographic data about the generated images. This allowed me to collect information on what the model perceives a janitor and a CEO to look like. I ended up generating around 30 images for each of the 10 occupations, resulting in a total of about 300 images.}'' In contrast, non-developer auditors preferred smaller scales typical of their daily work but would analyze a larger image corpus if the initial images had issues or did not meet their requirements. As P3 noted, ``\textit{I often focus on the image quality first... My  goals would be generating [ideal] images rather than testing the model.}''

\textbf{Auditing practices.} Generally, the most common method for auditing was described as ``just looking at the images''. As P4 noted, ``\textit{Because usually just like simply looking at them, you can tell [the issues].}'' However, participants acknowledged the time-consuming nature of examining images at scale. Specifically, P1 and P4 utilized generative AI tools with visual understanding capabilities, such as GPT-4 and LLaVA~\cite{llava}, to gain observations or insights. P1 described his workflow: \textit{``I use the visual assistant AI called LLaVA to examine images a lot... If you just create a pipeline where you create a bunch of images and then send all these images to the visual assistant AI, you can extract things like demographics on the images or just any random thing.''}  Nonetheless, these tools presented challenges, including inaccuracies or lackluster results (as noted by P4) and visual anomalies—like distorted human hands—that could only be identified by human reviewers (observed by P1).  P4 specifically mentioned: \textit{``So it's not 100\% accurate. That's the biggest challenge... it only works about like 90\% of the time.''}

\textbf{Auditing criteria.} The study shows that ViperaBase helped users identify unnoticed abnormalities compared to Baseline. Participants explained that this was because the additional scene graph serves as a ``check-list'', enabling them to examine the images in a disciplined manner, compared to merely relying on their experience and intuition in their usual workflow.  P2 elaborated on the benefits: \textit{``I think it's good at just like telling you what is in the image. There's a lot of things that you don't really notice when you're just staring at the image, staring at 20 images total. But when you get an AI to analyze every single image and compile all the data, you can really get a sense of what the model is printing out.''}  Moreover, all of them agreed that the node size encoding was helpful in uncovering minority objects, while comparing the sizes between adjacent nodes was also likely to yield interesting insights. As P3 noted: \textit{``So the fact that like the size of the circles corresponding to the amount of that item inside the image gives you a really good idea of the content of the images... So I think just like the most helpful thing is just the size of the circle.''} However, three participants mentioned that the dense scene graph was too time-consuming to go through, and they hoped to prune or customize the graph, such as cutting out nodes with very few data points (P1, P5) or adding evaluation criteria related to the keywords in their prompts (P3).


\subsection{Design goals}
Based on the findings and feedback from the design study, we have derived the following design goals (DGs):

\textbf{DG1: Leverage intuitive and interactive visual aids for structured, customizable auditing and effective result sensemaking. } According to the formative study, the visual aids, including the scene graph and the bar charts, might bring cognitive load to users, despite their effectiveness in inspiring audit directions and helping users understand the images. According to the user feedback, we infer two crucial reasons that account for this. First, the graphical layout and size of the scene graph may overwhelm users, causing confusion about where to focus during the analysis. The other 
was the lack of interactiveness in the visual aids: the users were forced to reactively check the suggested criteria instead of proactively creating them, limiting their creativity and activeness. Moreover, it is hard for users to associate the statistics back with the images due to the lack of coordinated views. Therefore, Vipera should grant users full agency through diverse interactions while allowing them to navigate between visual cues and data at any time. \looseness=-1

\textbf{DG2: Enable intuitive comparison of how the distribution of large-scale image outputs changes across prompts.} Comparing different images has been demonstrated as a rich source of inspiration for auditing in prior literature~\cite{everydayaudit, deng2025weaudit}. Likewise, we hypothesize that comparing the distributions and prompts may similarly provide abundant inspiration for users. To support this, Vipera should facilitate easy comparison of prompts and image distributions while highlighting the differences through proper visual assistance.

\textbf{DG3: Promote divergent thinking by highlighting image details and potential prompts that auditors might have overlooked.} Our formative study revealed that auditors often rely on intuition and personal experience, which can lead to confirmation bias and a failure to explore the model's ``\textit{unknown unknowns}''\cite{unknown-unknowns}. This tendency limits the scope of an audit to familiar issues. To break this pattern, Vipera should proactively surface unexpected or novel avenues for investigation. This involves drawing attention to subtle image differences that human reviewers might miss and suggesting new prompts that diverge from the user’s current line of inquiry, thereby encouraging a more comprehensive and less predictable auditing process.

\textbf{DG4: Integrating the LLM-powered guidance with visual guidance.} While LLM-powered suggestions are effective for inspiration, they can feel abstract or disconnected from the data if presented in isolation. Conversely, visual aids provide concrete evidence but may lack context or actionable next steps. To create a seamless and trustworthy workflow, Vipera should tightly couple these two modalities. This means that LLM-driven guidance (e.g., a suggested criterion) should be directly linked to the visual guidance (e.g., the scene graph), while the visual guidance characterizes the user's current state of auditing and provides contexts for the generation of targeted and visually explainable guidance. This multimodal loop aims to ground the AI's reasoning, making its suggestions more transparent and justifiable, while also making the visual data more actionable for the user.

\textbf{DG5: Assist individual auditors in documenting, synthesizing, communicating, and acting upon their audit findings.} The ultimate goal of an audit is to produce a coherent, evidence-backed report that can inform stakeholders and drive action. The process of documenting insights while simultaneously conducting an exploratory analysis is cognitively demanding and often disjointed. To address this, Vipera should integrate documentation directly into the analytical interface. This involves providing tools for auditors to seamlessly capture their thoughts, bookmark key visual evidence (such as charts or specific images), and structure their findings as they emerge, ultimately streamlining the transition from exploration to a communicable, actionable report.

\section{The Vipera System}
\begin{figure*}
  \centering
  \includegraphics[width=0.78\linewidth]{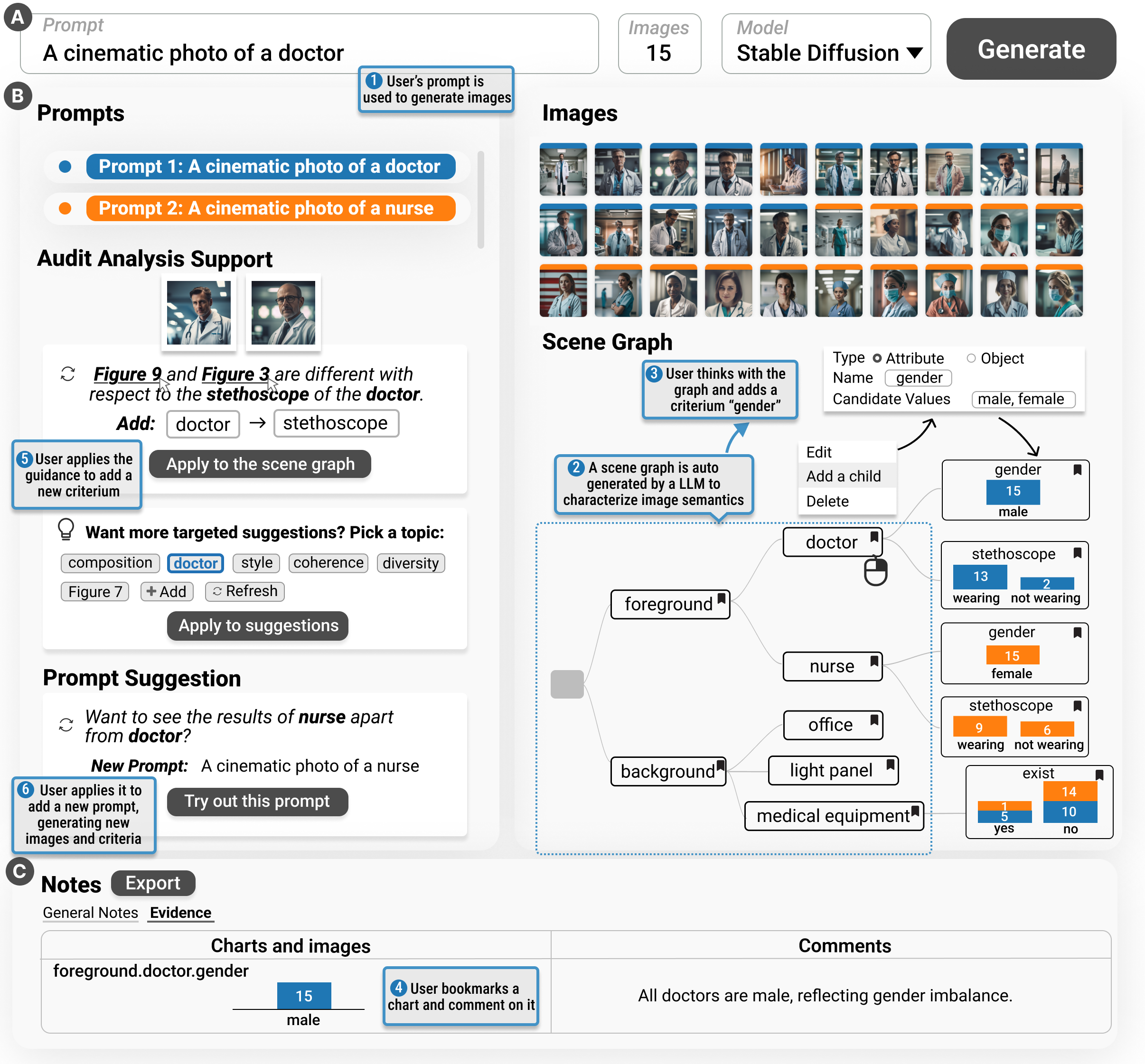}
  \caption{The Vipera interface. (A) The input box for creating prompts and specifying the number of images. (B) The analysis view for interactive auditing. Left: The prompts and AI-powered suggestions of auditing criteria and prompts. Right: The generated images and an interactive scene graph summarizing the image semantics to inspire users and guide the auditing process. (C) The note view for composing an auditing report. }
  \Description{Figure 1: The interface displays a tool for auditing generative text-to-image models. On the top (A), there is an input view which includes an input box for prompts, a numerical input box for the number of images, a selector for the model to be audited, and a "Generate" button for starting the image generation. Right below (A) is the analysis view (B). On the left side, a text box shows the prompt "A cinematic photo of a doctor," with options for audit analysis support with optional keywords for targeted suggestion generation and prompt suggestions. To the right, the images generated are placed here, annotated with different colors corresponding to the prompts. There is also a scene graph outlines the relationships between elements, including nodes such as "doctor", "nurse", "office", "medical equipment, and so on." Finally, there is a note view at the bottom, including a "General Notes" tab for overall insights and a "Evidence" tab for bookmarked charts and images. For instance, the current notes in the Evidence tab include a bar chart about doctors' genders and a comment pointing out the gender bias.}
  \label{fig:system}
\end{figure*}

This section presents the design and implementation details of Vipera. As shown in ~\autoref{fig:system}, three major components are included: an input view for creating prompts and generating images (A), an analysis view (B) for customizing the auditing process, seeking guidance, and analyzing the results, and a note view (C) for report generation. 

\subsection{Interface Design}
\subsubsection{Input view} 
The input view incorporates an input box for writing prompts, a numerical input box for specifying the number of images to be generated, and a model selector for choosing the model to be audited. The user can revisit this view at any time during the auditing to create new prompts or images. 

\subsubsection{Analysis view}
The analysis view consists of four subviews: prompts, images, scene graph, and audit suggestions.

\textbf{Prompts.} The prompts view displays all prompts created by the user, each encoded by a different color. By default, the results of all prompts will be displayed and analyzed. 

\begin{figure}[t]
  \centering
  \includegraphics[width=\linewidth]{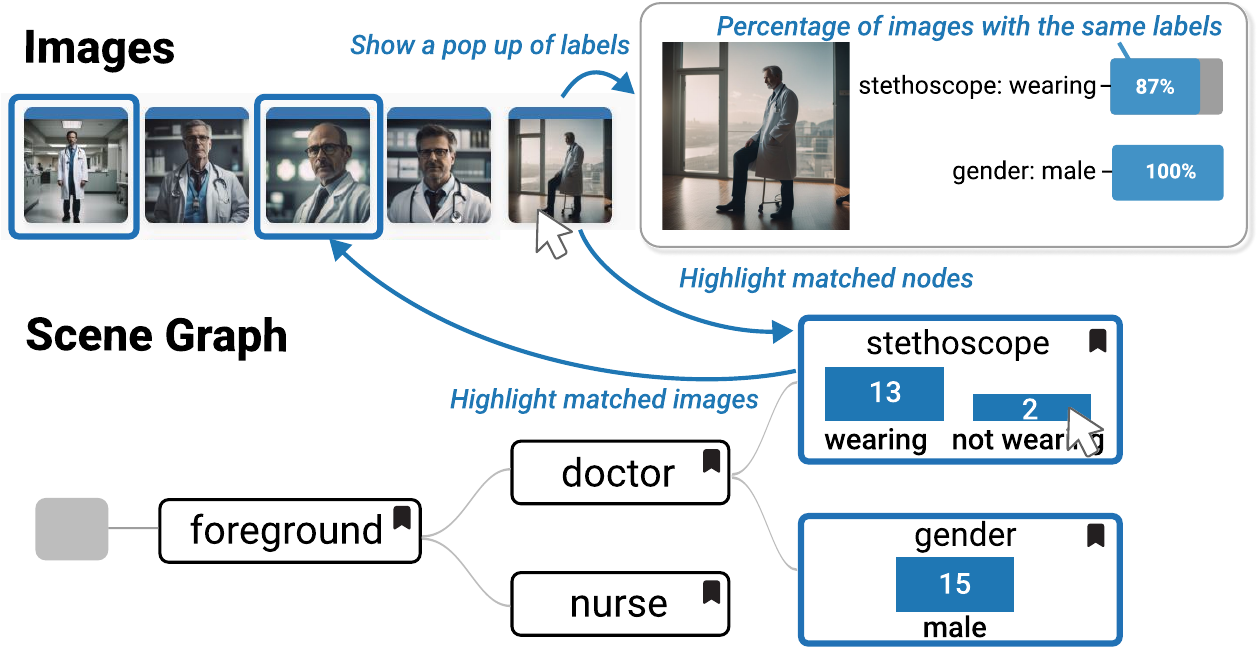}
  \caption{View coordinations between the image view and the scene graph. When the user hovers over an image, a pop-up will appear showing its labels, and the relevant attribute nodes in the scene graph will be highlighted. When the user hovers on a bar in an embedded bar chart in the attribute node, the corresponding images will be highlighted in the image view.}
  \Description{The above half of the image shows several images generated by the T2I model, while the bottom half shows the scene graph. The user hovers on an image, and a pop-up appears, showing the labels "stethoscope: wearing" and "gender: male", with a bar chart showing that 87\% and 100\% percent of images share the same labels in these criteria, respectively. Meanwhile, the nodes "stethoscope" and "gender" in the scene graph are highlighted. When the user hovers on the bar of "not wearing" of the bar chart embedded into the node "stethoscope", the images corresponding to this bar are highlighted in the image view.}
  \label{fig:coordination}
\end{figure}

\textbf{Images.} The image view shows the thumbnails of the generated images, with the border color indicating the corresponding prompt. Users can click on the images to see the full-sized version, where a bookmark button is provided for saving images of interest for later report generation. In addition, they can hover on the image to see the labels for the auditing criteria defined in the scene graph, and meanwhile, the associated nodes in the graph will be highlighted in the scene graph (\autoref{fig:coordination}). When they notice inaccurate labels, they are also allowed to edit the labels by right-clicking on the image, selecting the "Edit" option in the subsequent menu to enter a modal, and manually editing the labels within it. 

\textbf{Scene graph.} The scene graph serves as both a visual summary of the semantic contents within the images and a visual cue for users to organize their auditing (\textbf{DG1}). Objects in the scene graph are represented as nodes and organized in a tree layout based on their semantic relationship. The tree layout is used to replace the graphical one in ViperaBase for better intuitiveness (\textbf{DG1}), helping users organize their thoughts and navigate in the auditing space. 

To use the scene graph for auditing, users can add nodes to it by right-clicking on existing nodes and selecting the "add" option in the context menu to add a node. A user can specify the following information when adding a node:
\begin{itemize}
    \item \textbf{Node type:} Vipera supports two types of nodes: \textit{object} nodes and \textit{attribute} nodes. Object nodes represent physical objects or concepts in the images, while attribute nodes are leaf nodes that represent specific properties or characteristics of their parent object node. For example, non-leaf nodes in ~\autoref{fig:system} such as \textit{foreground} and \textit{doctor} are object nodes that describe the image semantics, while leaf nodes like \textit{gender} are attribute nodes that can used to evaluate their parent nodes.
    \item \textbf{Node name:} The name of the node to be added.
    \item \textbf{Scope:} The scope of each node defines the prompts or images to which the current evaluation criteria (for attribute nodes) or the criteria of descendant attribute nodes (for object nodes) apply. This will help prevent evaluating all images against all criteria, which could be overwhelming and lead to intent misalignment (e.g., evaluating the nurse images with the criteria ``doctor's gender''). Users can choose to apply the criteria to all prompts, specific prompts, all images, or specific images. By default, all images and prompts will be selected.
    \item \textbf{Scope type:} Since Vipera supports iterative creation of prompts and auditing criteria, the scope of nodes in the scene graph can be dynamic to accommodate the prompts and images to come. Given this, Vipera supports two scope types: \textit{fixed} and \textit{auto-extended}. If an auto-extended scope is specified, the criteria will be applied to newly generated images, which is not the case for a fixed scope. 
    \item \textbf{Candidate values (optional):} To prevent the LLM from generating overly diversified labels for the images that hardly lead to meaningful distinctions, Vipera allows users to specify a list of predefined candidate label values for classification (e.g., \textit{male} and \textit{female} for the attribute node \textit{gender}). If provided, the LLM will be constrained to choose from this list when labeling images. 
\end{itemize}

When an attribute node is added, the node will be used as a criterion for a labeler LLM to label the images. The result labels will be visualized in stacked bar charts with colors indicating the corresponding prompts to facilitate comparison (\textbf{DG2}). Users can hover on the bars to navigate back to the matched images highlighted in the image view (\autoref{fig:coordination}).  Additionally, in case the labeler itself makes mistakes, users can edit the information of a node or choose to relabel the images based on an attribute node via the context menu. A relabeling will also be triggered when the user modifies the candidate values or the scope. In this case, two options are provided: relabeling all images or only relabeling images that are affected (i.e., contain a label that is not within the candidate values or not in the scope). The user can change the option in the relabeling modal.

\textbf{Suggestions.} Informed by \textbf{DG3}, Vipera provides two classes of auditing guidance, \textit{audit analysis support} and \textit{prompt suggestion}, to inspire users of new auditing criteria and prompts, respectively. 

\begin{itemize}
    \item \textbf{Audit analysis support.} This module highlights two images with notable differences and suggests auditing additional auditing criteria as nodes. To mitigate intent misalignment caused by random selection, Vipera enables users to choose from a diverse set of LLM-generated keywords or create their own custom keywords for more targeted suggestions.  These keywords are injected into the prompt, and suggestions are presented to the user only when the LLM’s self-reported confidence score exceeds a predefined threshold after several iterations, indicating a reliable identification of topic-relevant differences between the images.

    \item  \textbf{Prompt suggestion.} This module promotes divergent thinking by suggesting insightful prompts, replacing words or phrases in existing prompts. Upon users deciding to try out the prompt, Vipera will duplicate the existing auditing criteria while generating new images, allowing for an effective comparison of results between different prompts. For instance, when the user applies the suggestion replacing the word ``\textit{doctor}'' with ``\textit{nurse}'' in the prompt ``\textit{a cinematic photo of a doctor}'', Vipera will automatically add an object node ``\textit{nurse}'' to the sibling of ``\textit{doctor}'' while duplicating the descendent nodes of ``\textit{doctor}'' in this new branch. Such a transformation makes sure that the labeling results of newly generated images for nurses will be displayed in the correct branch.
\end{itemize}

\begin{figure}[t]
  \centering
  \includegraphics[width=\linewidth]{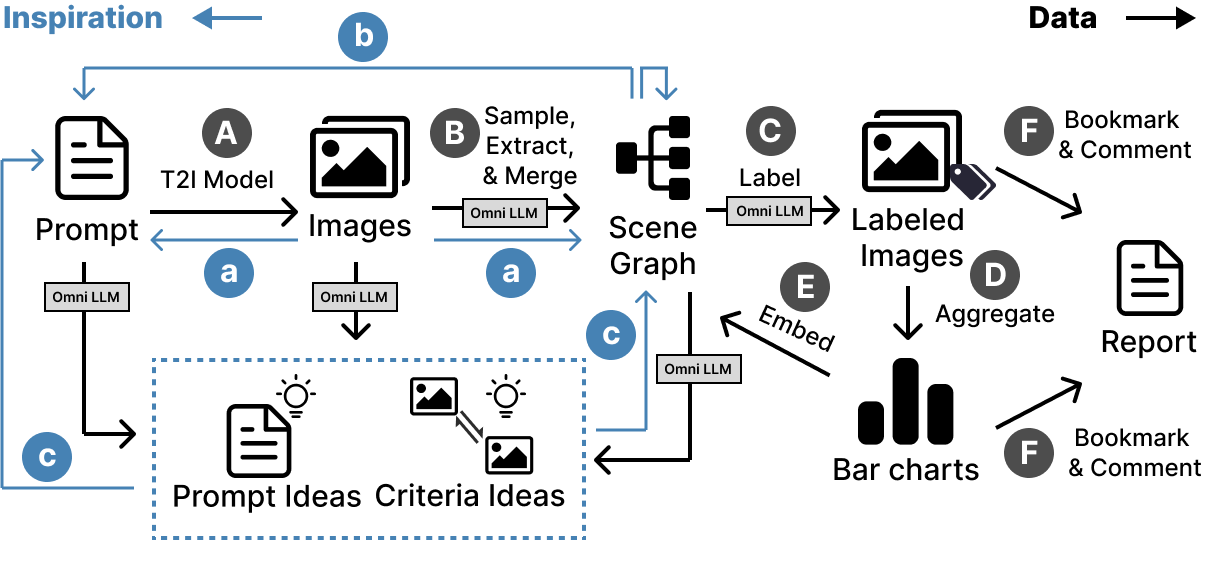}
  \caption{Vipera's technical pipeline. Black edges indicate the data flow: (A) The prompt from the user is fed to the T2I model to generate images. (B) A scene graph is generated from the images (see \autoref{fig:scene_graph_gen} for details). (C) Upon adding an attribute node to the scene graph, the images will be labeled according to the specified criteria. (D) The labels will be aggregated and visualized as a stacked bar chart. (E) The charts will be embedded into the attribute nodes of the scene graph for rendering. (F) The images and charts will be added to the auditing report upon being bookmarked by the user, with the user's comments attached to them. Blue edges indicate the inspiration flow for iterative auditing: Prompts and auditing criteria can be inspired by inspecting the images (a), thinking with the scene graph (b), and applying the LLM-driven guidance (c).}
  \Description{The flowchart illustrates a process for generating and analyzing images using the T2I model. It begins with a prompt that leads to the creation of images. These images are then sampled and extracted to form a scene graph. The scene graph is labeled to produce labeled images, which can be aggregated and visualized in bar charts that will later be embedded into the scene graph. Additionally, there is a section for prompt ideas and comparisons, generated by the LLMs using the prompts, images, and scene graph. Meanwhile, there is also an inspiration flow complementary to the above data flow, where users can get auditing inspirations on prompts and criteria from inspecting images, thinking with the scene graph (or checking the charts embedded), or applying the LLM-powered suggestions, and consequently create more prompts and criteria for iterative auditing. The entire process is driven by an inspiration flow (in blue arrows), which flows from right to left in the diagram. Another data flow (in black arrows, flowing from left to right) is incorporated to generate content that powers the inspiration flow.}
  \label{fig:workflow}
\end{figure}

\subsubsection{Note view} \label{vipera:note}
To facilitate the authoring of auditing reports for individual auditors (\textbf{DG5}), a note view is incorporated, which consists of two tabs: a tab of \textit{General Notes} that includes an input box for noting down the overall insights, and a tab of \textit{Evidence} that shows the bookmarked charts and images. Detailed user comments can be appended to specific charts or images as fine-grained insights. Moreover, Vipera features LLM-powered auto-completion when users write in this view, further enhancing auditing efficiency. The prompts,  bookmarked items, and existing notes will be injected into an LLM to generate the completion, which can be easily applied when users press the \textit{tab} key on the keyboard.

\subsection{User workflow and technical pipeline}
\begin{figure}[t]
  \centering
  \includegraphics[width=\linewidth]{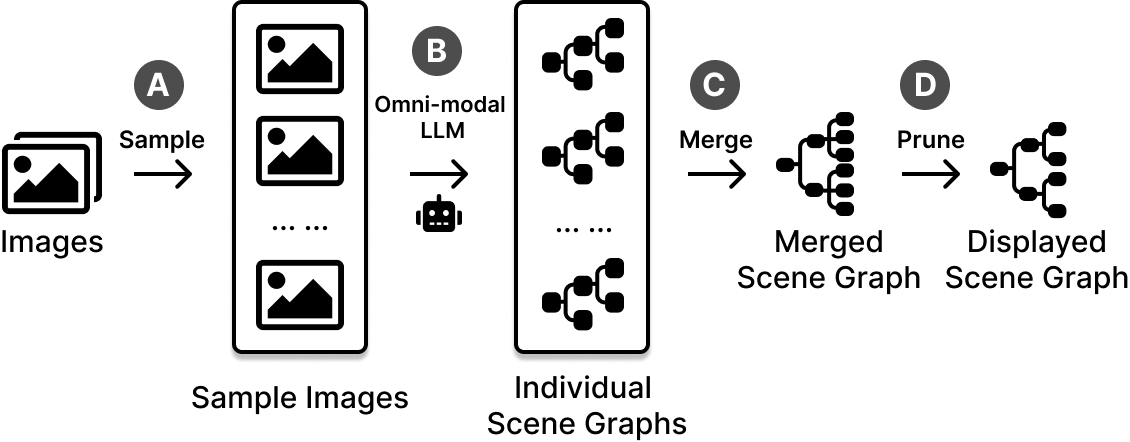}
  \caption{The detailed technical pipeline for generating the initial scene graph. (A) A random subset is sampled from the images. (B) For each image, an individual scene graph is generated through an omni-modal LLM. (C) The result scene graphs are merged into one. (D) The result is pruned by keeping a maximum of five leaf nodes.}
  \Description{Same as the caption of this figure.}
  \label{fig:scene_graph_gen}
\end{figure}

Vipera has been developed and deployed as a web-based application. \autoref{fig:workflow} shows the workflow of Vipera\footnote{All system prompts used in the pipeline are provided in the appendix.}. After generating images from a user's prompt, Vipera initiates the scene graph construction process, which is detailed in \autoref{fig:scene_graph_gen}. It starts by randomly sampling four images (or all, if fewer than four are available) to serve as a basis for the initial scene graph (\autoref{fig:scene_graph_gen}A). We chose this small sample size based on internal piloting, which indicated it struck an effective balance between capturing a diversity of concepts and ensuring low latency for the user, thus prioritizing usability and speed (DG1). 

For each sampled image, Vipera prompts an omni-modal LLM (Gemini 2.5 Flash in our implementation) to extract a tree-based scene graph that captures the image's semantics (\autoref{fig:scene_graph_gen}B). The LLM is instructed to identify key physical objects and notable features and organize them into a tree structure. To provide a consistent high-level organization, we constrain the first level of the tree to ``\textit{foreground}'' and ``\textit{background}'' nodes in our study, which is an intuitive and widely applicable distinction. The resulting scene graphs from the four images are then merged, with their leaf nodes combined (\autoref{fig:scene_graph_gen}C). To avoid overwhelming the user (\textbf{DG1}), we ensure the scene graph's size is approximately seven nodes as informed by the Miller's Law. Thus, a maximum of five\footnote{Our informal sensitivity checks suggested that minor variations in the scene graph parameters, such as sampling 3 or 6 images, or adjusting the node limit, did not qualitatively alter the conceptual breadth of the graph.} leaf nodes are randomly selected for the final aggregated scene graph (\autoref{fig:scene_graph_gen}D), ensuring fairness and giving rarer but potentially insightful concepts an equal chance to appear in the initial scaffold. Note that the initial graph is not an exhaustive summary but rather a lightweight, interactive starting point. Subsequent user refinement during the auditing process inherently mitigates the impact of the initial random sampling.

With the scene graph serving as a visual summary, users can add object or attribute nodes, where attribute nodes will serve as auditing criteria. When an attribute node is added, Vipera begins to evaluate and label the images using this criterion (\autoref{fig:workflow}C). First, it updates the scene graph with the new node and ensures that all images within the new node's scope are also included in the scope of its ancestor nodes up to the root. Then, it extracts a \textit{partial graph schema}, which is a subset of the scene graph containing the new node and its path to the root. This schema is used to prompt an omni-modal LLM (GPT-5-mini in our implementation) to generate a label for each image within the specified scope. If the user has defined candidate values for the attribute, the LLM is constrained to choose from that list. The resulting labels are then aggregated, and their distribution is visualized as a stacked bar chart (\autoref{fig:workflow}D) embedded into the added node using the D3.js library\footnote{https://d3js.org/} (\autoref{fig:workflow}E). Furthermore, users can bookmark the images and the charts within scene graph nodes at any time during auditing and comment about them in the note view for an auditing report (\autoref{fig:workflow}F).

Additionally, users can benefit from diverse sources of inspiration for iterative exploration of prompts and auditing criteria in Vipera, as illustrated by the blue arrows in \autoref{fig:workflow}. Specifically, they may inspect the images carefully to get insights and new inspirations for prompts and criteria (\autoref{fig:workflow}a). Such inspirations may also come from the scene graph, where the user either thinks with its structure about potential nodes to be added or examine the embedded charts for insights (\autoref{fig:workflow}b). Finally, they may adopt the LLM-powered suggestions at any time for new auditing ideas (\autoref{fig:workflow}c).  Such an inspiration loop is built upon the artifacts from the data loop and  serves as the engine for the user workflow, promoting iterative, mixed-initative, and comprehensive auditing.

\subsection{Usage Scenario}
Now we illustrate Vipera's workflow using the story of Bob, an auditor tasked with evaluating the Stable Diffusion model for potential biases. \looseness=-1

To begin his audit, Bob enters the prompt ``\textit{A cinematic photo of a doctor}'' into the input view and generates 15 images (\autoref{fig:system}, step 1). Vipera populates the analysis view with the prompt and images, while displaying a corresponding scene graph that characterizes the semantics of the images (\autoref{fig:system}, step 2). The graph organizes the content into \textit{foreground} (e.g., \textit{doctor}, \textit{nurse}) and \textit{background} (e.g., \textit{office}, \textit{medical equipment}) nodes. 

With the scene graph as an overview of the images' contents, Bob organized his thoughts and decided to start from the \textit{doctor} node. He right-clicked on this node and chose to add a child in the context menu (\autoref{fig:system}, step 3). The new node was an attribute node named \textit{gender} with candidate values \textit{male} and \textit{female}. Vipera automatically labeled the images based on the gender of the doctor and visualized the results in the bar chart embedded in the new node. Realizing all doctors were male from this chart, Bob bookmarked it as evidence and documented his insights in the notes view (\autoref{fig:system}, step 4). 

Afterwards, Bob felt too tired to come up with new auditing ideas and decided to refer to the AI suggestions. He first examined the Audit Analysis Support, which highlighted two images with a key difference and suggested adding a \textit{stethoscope} attribute to the \textit{doctor} node. Bob found the suggestion insightful and applied the suggestion (\autoref{fig:system}, step 5). The resulting chart showed that some doctors were not wearing a stethoscope. To verify this, he hovered over the corresponding bar, which highlighted the specific images in the image view (\autoref{fig:coordination}), allowing for rapid, coordinated sensemaking.

Next, Bob noticed the \textbf{Prompt Suggestion} recommending he replace ``doctor'' with ``nurse'' to explore a related profession. He accepted the prompt, and Vipera generated a new set of images (\autoref{fig:system}, step 6). Crucially, the system also added a new \textit{nurse} branch to the scene graph and automatically duplicated the existing criteria (\textit{gender} and \textit{stethoscope}) under it. With Vipera's color-coded bar charts, Bob compared the results of these two prompts on the scene graph easily and acquired new insights into occupational stereotypes.  From this point, Bob can continue his audit through multiple iterations of structured exploration and comparison.
\section{User Study}
We have conducted a user study to evaluate the usability and effectiveness of Vipera. Our main goal was to understand how Vipera's core features—the Scene Graph and AI-powered suggestions—influence the process and outcomes of different AI auditing tasks. We were also interested in the usage patterns revealed from the user's auditing process.

\subsection{Study Design and Methodology}
\textbf{Participants.} We recruited 24 participants (P1-P24, 14 males and 10 females). Among them, 20 were general student auditors recruited from our institution (4 undergraduates, 7 Master's, and 9 Ph.D students) who reported an average familiarity with AI auditing (M=3.04/5, SD=1.16) and use T2I models with an average frequency (M=2.92/5, SD=0.929). The remaining 4 participants were expert auditors recruited from both academia and industry who had rich experience and knowledge in AI auditing. The ages of the participants ranged from 20-29, with an average of 23.9.

\textbf{Baseline.} There were four systems (A-D) used in the study, including three baseline systems adapted from Vipera (A-C) and the original Vipera system (D).

\begin{figure}[t]
  \centering
  \includegraphics[width=0.8\linewidth]{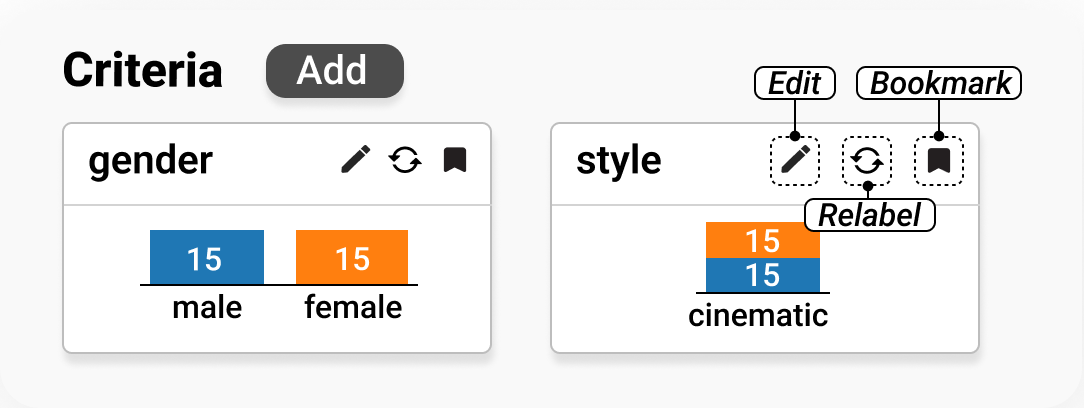}
  \caption{The additional Criteria View added to Systems A and C. Users are allowed to add auditing criteria, and the results will be visualized as bar charts in the respective cards. Each criterion card includes three buttons for editing the criterion, relabeling the images, and bookmarking the criterion, respectively.}
  \Description{In this view, there is a view title "Criteria" and an  "Add" button next to it for users to add an auditing criterion. There are currently two cards in this view, one for each criterium: a "gender" card embedding a bar chart showing 15 male (blue) and 15 female (orange) in parallel, and a "style" card embedding a bar chart showing 30 images under the "cinematic" label (15 blue and 15 orange ones, visualized as a stacked bar chart. On the top right of each card, there are three buttons: edit, relabel, and bookmark. }
  \label{fig:criteria}
\end{figure}

\begin{itemize}
    \item \textit{System A}: The system was adapted from Vipera by removing both the AI auditing support and the scene graph. To equip the system with minimum auditing capabilities, we added an additional \textit{criteria view} to this system allowing criteria creation and result visualization, as depicted in \autoref{fig:criteria}. User can add auditing criteria, set candidate values and scope for labeling as in Vipera.  

    \item \textit{System B}: The system was adapted from Vipera by merely removing the AI auditing support.

    \item \textit{System C}: The system was adapted from Vipera by removing the scene graph and replacing it with the same criteria view as in System A.
\end{itemize}

\textbf{Tasks.} Participants were asked to audit the Stable Diffusion XL model.  They were assigned one of four initial prompts designed to elicit a range of potential biases and quality issues: (a) 
\textit{A couple on their wedding day}, (b) \textit{A family having a picnic in the park}, (c) \textit{Worldwide athletes in the Olympic Games}, or (d) \textit{An award-winning chef preparing a gourmet meal}. These prompts are able to cover diverse topics and image features, such as individuals (a, d), crowd (b, c), country (c), occupation bias (a, c), cultural issues and potential bias (c, d).
Participants were instructed to explore the model's behavior freely from this starting point.

\begin{figure*}[t]
  \centering
  \includegraphics[width=\linewidth]{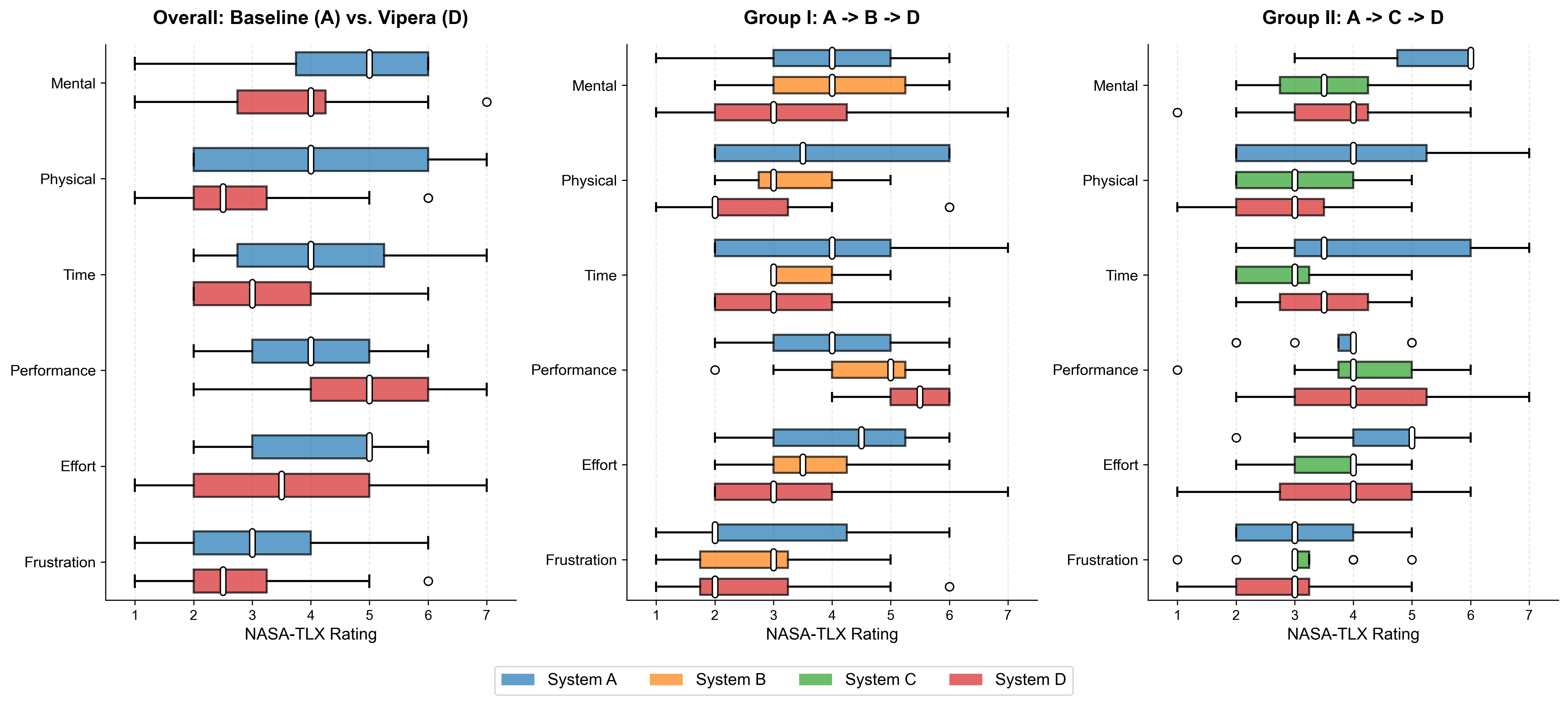}
  \caption{Participants' ratings on the NASA-TLX scale. The first subplot shows a rating comparison of all participants on the Baseline (A) and Vipera (D), while the other two subplots show a detailed rating comparison for the three systems used by participants in each group. The white marks on the boxes indicate the medians. }
  \Description{In the first subplot, the box plot shows that participants perceived less workload and better performance. The second subplot shows that participants perceived a decreasing level of workload from A, B, to D in Mental, Physical, Time, Effort and an increasing level of performance. For Frustration, the median of A is similar to D, and both are smaller than the median of B. The third subplot shows that participants perceived similar workloads in Time and Frustration and similar Performance. The Mental,  Physical, and Effort workload of C and D is smaller than A, and there is no significant difference between C and D. }
  \label{fig:nasa}
\end{figure*}

\textbf{Procedure.} Each session lasted approximately 90 minutes. After providing informed consent, participants completed a demographic questionnaire. They were then equally divided into two groups (Group I and Group II), with the distribution of general and expert auditors kept consistent across both groups. The study employed a mixed design where participants in Group I used systems A, B, and D, while those in Group II used systems A, C, and D. This progressive design allows us to observe how their auditing process and performance evolve as new capabilities are introduced. While a counterbalanced design in system order may mitigate the learning effect, it may introduce a more problematic contrast effect where participants evaluate the simpler system based on its missing features, rather than its intrinsic utility.

Before using each system, they were provided a tutorial on the system, followed by a warm-up session of up to 5 minutes to familiarize themselves with the interface. Subsequently, they were given 15 minutes to perform an auditing task, with the goal of acquiring as many insights as possible. To balance between controlled comparison and free exploration, all participants began with their assigned prompt but were free to write subsequent prompts on the same topic.
Each starting prompt was assigned to an equal number of participants. They were allowed to use the same prompts across systems so we could more directly attribute differences in user strategy, efficiency, and audit outcomes to the system itself, clarifying each component's impact on the user's workflow for a consistent task.

Participants were also instructed to note down their insights in phrases or short sentences as soon as they got them to eventually form an auditing report after using each system.
We employed the think-aloud protocol~\cite{think-aloud} to capture their reasoning process. After completing the tasks with all three systems, participants filled out a questionnaire that included NASA-TLX scales for each system and several 7-point Likert scales to rate the usefulness of system components. The study concluded with a semi-structured interview to gather qualitative feedback. Participants received a \$10 gift card as compensation for their time. All sessions were both screen- and audio-recorded. The study proposal has been approved by the IRB committee of our institution.

\textbf{Data collected and analysis.} We collected multiple forms of data for each participant: three auditing reports (one for each system used), pre- and post-study questionnaires, system interaction logs, and full session recordings. Our analysis approach was twofold. Quantitative data from the questionnaire responses and system interaction logs were analyzed using statistical methods to assess performance and usability. Qualitative data from the auditing reports and transcripts from the think-aloud protocol and semi-structured interviews were analyzed using thematic analysis~\cite{thematic-analysis} to identify user strategies, reasoning patterns, and key feedback themes.

\section{Results}

In this section, we share the results from our controlled experiments. In particular, analysis of the NASA-TLX questionnaire responses revealed that participants reported lower workload and improved performance when using Vipera—especially with the AI auditing support—compared to the baseline. Auditing logs indicate that Vipera assisted participants in creating more auditing criteria through interaction with AI support and visual guidance. Our interviews with practitioners also revealed nuanced trade-offs between scaffolding systematic auditing while managing cognitive demand, personalizing recommendations, and leveraging resources tailored to individual auditors.

\subsection{Mixed-method Analysis of Questionnaires and Audit Logs} \label{result:mixed-method}

\begin{figure*}[t]
  \centering
  \includegraphics[width=\linewidth]{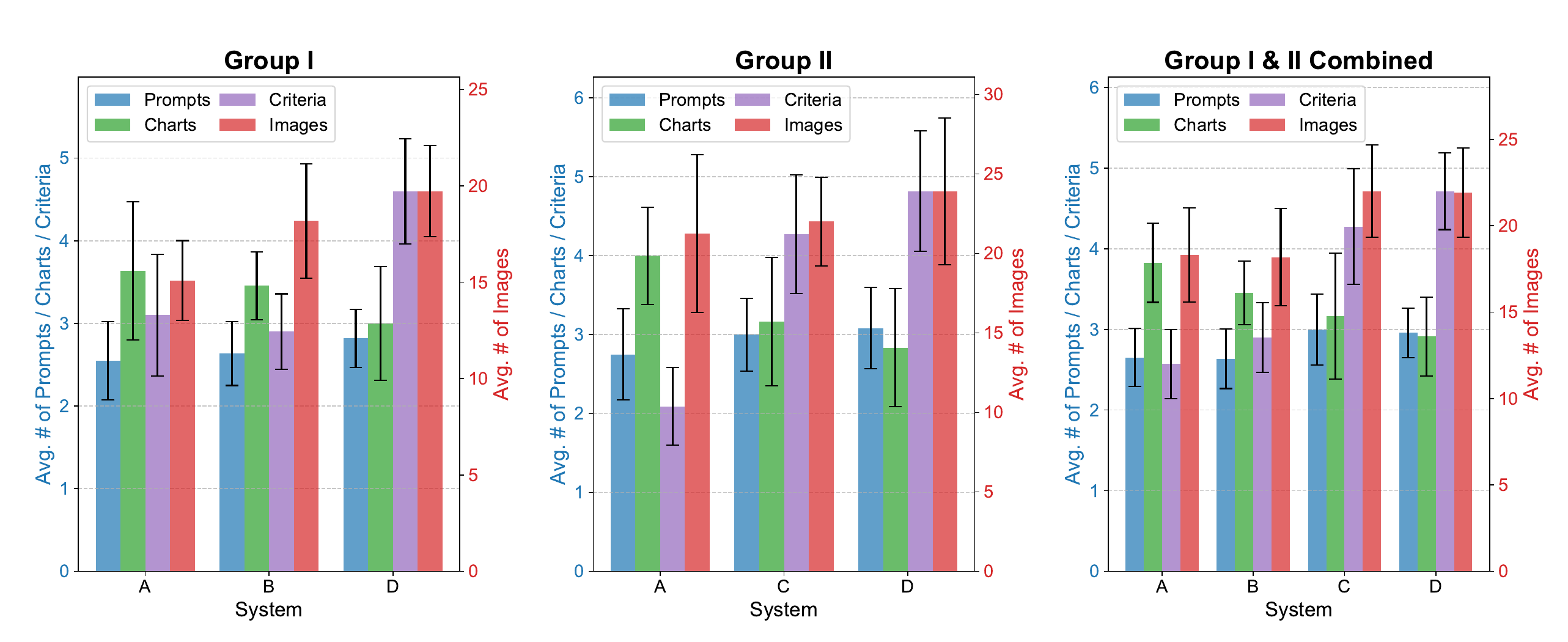}
  \caption{The number of prompts, images, used criteria, and bookmarked charts or images in each system used by the participants. Error bars indicate standard error of the mean.}
  \Description{There are three subplots in this figure. The first subplot shows the number of prompts, images, used criteria, and bookmarked charts or images in each system used by the participants of Group I (i.e., the systems are A, B, and D). For each system on the x-axis, there are four bar charts indicating the number of prompts, charts, criteria, and images from left to right, each with an error bar. Similarly, the second subplot corresponds to participants in Group II (i.e., systems A, C, and D) with a similar layout, while the third subplot corresponds to all participants (i.e., systems A, B, C, and D). }
  \label{fig:log_line}
\end{figure*}

\subsubsection{\textbf{Questionnaires}}

The detailed distribution of participants' ratings on NASA-TLX scales is illustrated in \autoref{fig:nasa}. 
We analyzed ratings using paired t-tests, with p-values ($p$) corrected for multiple comparisons using the Holm–Bonferroni method ($p'$). Our analysis first compared the final Vipera system (D) against the baseline (A) for all participants, then examined the step-wise introduction of components within each group to understand their individual contributions.


Our primary finding is that Vipera (D) delivered a statistically significant improvement in participants' self-rated Performance compared to the baseline (A) ($t(23)$ = 2.685, $p$ = 0.0066, $p'$ = 0.0397). While participants also reported lower average workload with Vipera, particularly in Mental Demand ($t(23)$ = -2.055, $p$ = 0.0257, $p'$ = 0.0514) and Physical Demand ($t(23)$ = -2.061, $p$ = 0.0254, $p'$ = 0.0762), these reductions represented a positive trend but were not statistically significant after correction.

To understand the naunces of why a clear performance gain does not accompany a corresponding significant drop in overall workload, we analyzed the data from each experimental group. The results reveal that the two main components of Vipera have distinct and somewhat opposing effects on user workload.

For Group II, introducing the AI auditing support alone (System C) caused a dramatic and statistically robust reduction in Mental Demand ($t(11)$ = -5.063, $p$ = 0.0002, $p'$ = 0.0022) compared to the baseline. This powerful effect, which remained highly significant even after correcting for 12 planned comparisons, establishes the AI component as the primary driver of workload reduction in Vipera. Other workload dimensions like Effort ($p$ = 0.0086, $p'$ = 0.0515), Physical Demand ($p$ = 0.0380, $p'$=0.0912), and Temporal Demand ($p$ = 0.0310, $p'$ = 0.0931) also showed trends toward reduction.

In contrast, the experience of Group I participants, who received the scene graph first (System B), was more mixed. We observed trends towards improved Performance when adding the scene graph (A vs. B, $t(11)$ = 3.023, $p$ = 0.0058, $p'$ = 0.0696) or subsequently adding the AI support (B vs. D, $t(11)$ = 1.876, $p$ = 0.0437, $p'$ = 0.1748), although neither remained robust after correction. We hypothesize that this inconsistency may be due to varying degrees of benefit derived from the scene graph across participants. In fact, descriptive data shows that median Mental Demand was slightly higher for System B than for the baseline (A) or Vipera (D), suggesting the scene graph may introduce its own cognitive overhead. However, subsequent addition of the AI auditing support reduces workload and improves performance on average.

To summarize, the results indicate that the \textbf{AI auditing support can reduce users' mental workload}, while the \textbf{scene graph can contribute to auditing performance despite potential addition to the temporal demand}. Most participants attributed this extra temporal demand to the significant amount of time spent on exploring promising auditing criteria when the scene graph is present. A few participants also mentioned the fact that it is naturally difficult to find useful criteria and get insights in systems used at a later time when they have already explored the model extensively in the previous systems.

Finally, participants generally perceived all components within Vipera as helpful, with an average rating of 3.50 (SD=0.885) and 3.67 (SD=0.868) on the scene graph and AI auditing support, respectively. They also appreciated the automatic labeling of images (M=3.375, SD=0.970) and note-taking features (M=3.71, SD=1.083), indicating the overall high usability of Vipera interface.



\subsubsection{\textbf{Auditing logs}}\label{auditing-logs}

\begin{table*}[t]
  \centering
  \caption{Thematic analysis results of the participants' auditing insights. }
  \label{tab:insights}
  \small
  \begin{tabular}{@{} l l p{6.5cm} @{}}
    \toprule
    Category & Subcategory & Example quotes (participant ID) \\
    \midrule
    A. Quality & A1. Exquisiteness &
      ``The details are slightly lacking'' (P08); ``The picture looks beautiful'' (P16); ``Mosaics appearing'' (P19)\\
    & A2. Authenticity &
      ``The person looks like a dummy'' (P02); ``The dishes are floating'' (P12); ``People's faces look distorted'' (P05) \\
    & A3. Style &
      ``Prefer to see more realistic images but the majority are artistic'' (P21) \\
    & A4. Diversity &
      ``The characters are highly homogeneous'' (P19); ``Racial bias'' (P02); ``AI may discriminate unwealthy people'' (P13) \\
    \addlinespace
    B. Prompt Interpretation & B1. Alignment with the prompt &
      ``The content of the picture aligns with the 'worldwide' theme'' (P15); ``No 'athletes' in the image'' (P03) \\
    & B2. Alignment with common sense &
      ``There are six fingers in one hand'' (P08); ``People's mood is wrong; they should be happy on the wedding day'' (P05) \\
    & B3. Alignment with user understanding &
      ``Even if I added the word 'beautiful' to the prompt, the images were still ugly'' (P02); ``It misunderstood me'' (P03) \\
    \addlinespace
    C. Reasoning for LLM Behavior & C1. Prompt comparison &
      ``The pictures are better when more details are included in the prompt'' (P06) \\
    & C2. Express the tendency of the LLM &
      ``The chance of errors increases when more elements exist'' (P16); ``LLM didn't understand subjective keywords well'' (P02) \\
    & C3. Explain for the LLM behavior &
      ``I probably want a more diverse set of colors. Could be attributed to most of the training images having this color'' (P21) \\
    \addlinespace
    D. Critic on the System &  &
      ``AI gives me more options... But the recommended prompt may have little novelty'' (P14); ``The classification of images is problematic'' (P06) \\
    \bottomrule
  \end{tabular}
\end{table*}

We measured the number of prompts, generated images, used criteria, and bookmarked charts or images used by each user in each system, as illustrated in \autoref{fig:log_line}. Note that we use the number of bookmarks as a proxy to evaluate the number of insights despite potential misrepresentation, given that it is hard to define an atomic insight for calculation. 
The results show that the introduction of guidance had a notable effect on user behavior. Specifically, users created more auditing criteria in all guided systems (B, C, and D) compared to the baseline (System A), with the highest number of criteria being used in the full Vipera system (System D). A similar trend was observed for prompts and images, though the increase was primarily driven by the systems featuring AI auditing support (Systems C and D). In contrast, the addition of the scene graph alone (System B) did not lead to a noticeable increase in the number of prompts or images.
Conversely, we observed a consistent and significant decrease in the number of bookmarked charts across the guided systems. This effect was most pronounced in Systems C and D, suggesting that the presence of AI support may have shifted users' focus from documenting existing views to actively exploring new criteria and prompts. This interpretation aligns with participants' feedback in the interview, which indicated that more effort was spent on exploration when using Vipera's advanced features.

One interesting discovery is that the introduction of AI auditing support (System A vs. C; System B vs. D) consistently leads to an increase in the number of prompts, images, and criteria, as well as a consistent decrease in the number of bookmarked charts or images. This shows that auditing support powered by AI is beneficial and inspirational for most participants despite the extra effort in exploration. By contrast, the effect brought by the introduction of the scene graph (System A vs. B; System C vs. D) was more diverse and random, varying from person to person.

We also evaluated the auditing pattern and efficiency of participants. Overall, we found that in Systems C and D, respectively, 87.2\% and 78.4\% of the criteria were authored by AI, while only
61.8\% and 50.8\% of the prompts were generated by AI. This reveals that participants desired more agency and control in writing prompts than criteria. We also observed a high consistency in users' prompts: the average pairwise BERT~\cite{devlin2019bert} cosine similarity between all user-generated prompts within each session was 0.904 (on a 0–1 scale), showing that most participants intentionally utilized similar prompts to facilitate comparison. 
Meanwhile, we observed that  each prompt led to 1.2, 1.2, 0.9, and 0.8 bookmarks on average in Systems A-D, respectively, showing a decrease in auditing efficiency despite the broadened auditing scope. 

Notice that we also calculated the average number of prompts, images, criteria, and bookmarks for each starting prompt, but we didn't find any major and consistent impact of the starting prompt on these metrics.

\subsubsection{\textbf{Auditing reports}}

\begin{figure*}[t]
  \centering
  \includegraphics[width=\linewidth]{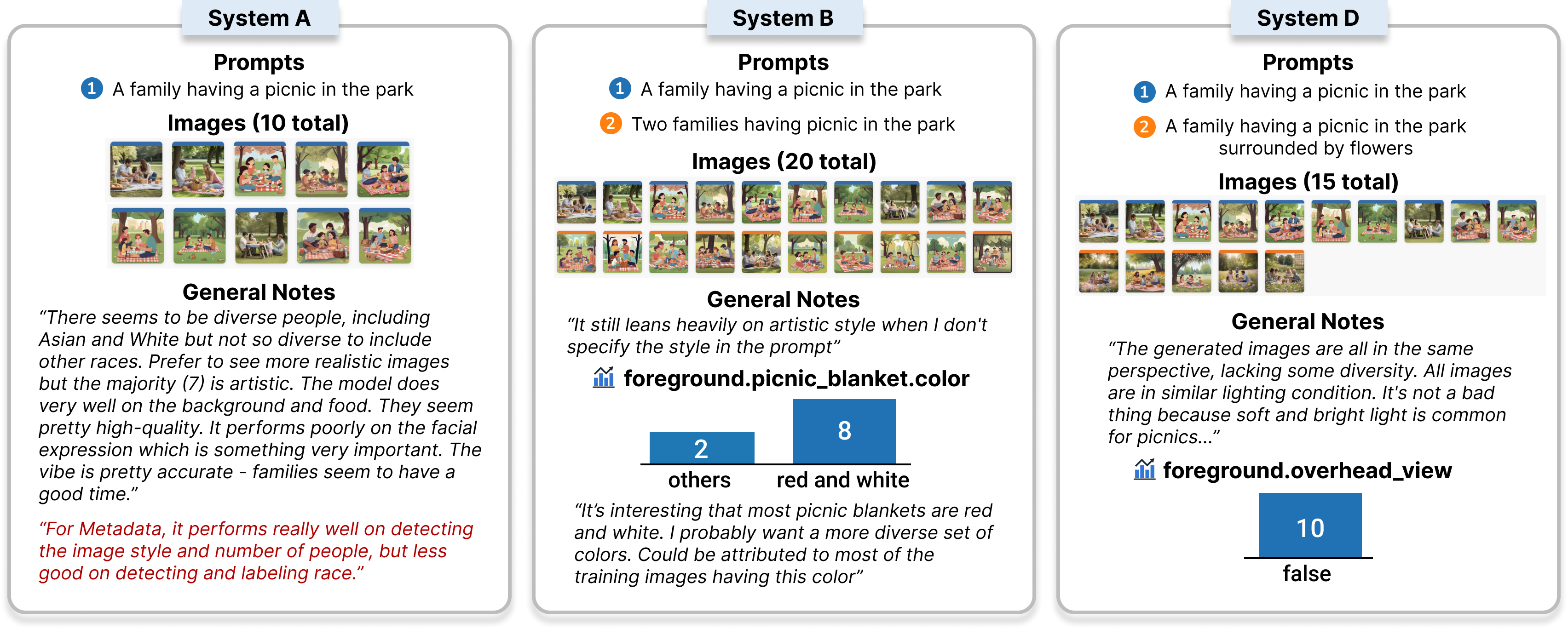}
  \caption{Example auditing reports from one participant, P21. Reports are partially truncated due to the length.}
  \Description{In the first report (System A), the participant used one prompt "A family having a picnic in the park" to generate 10 images, and wrote the following general notes: "There seems to be a diverse group of people, including Asian and White but not so diverse to include other races. It could also be attributed to the artistic style of the images, which make race hard to tell. Prefer to see more realistic images but the majority (7) is artistic. The model does very well on the background and food. They seem pretty high-quality. It performs poorly on the facial expression which is something very important. The vibe is pretty accurate - families seem to have a good time." In the second report (System B), the participant used two prompts to generate 10 images each: "A family having a picnic in the park" and "Two families having picnic in the park". The general notes was "It still leans heavily on artistic style when I don't specify the style in the prompt". Furthermore, a bar chart "foreground.picnic\_blanket.color" was cited with 8 "red and white" and 2 "others". The participant wrote, "It’s interesting that most picnic blankets are red and white. I probably want a more diverse set of colors. Could be attributed to most of the training images having this color". In the third report (System D), the participant used two prompts to generate 10 images and 5 images respectively: "A family having a picnic in the park" and "A family having picnic in the park surrounded by flowers". The general notes was "The generated images are all in the same perspective, lacking some diversity. All images are in similar lighting condition. It's not a bad thing because soft and bright light is common for picnics...". A bar chart "foreground.overhead_view" was also cited showing that all images were not taken from overhead.}
  \label{fig:example}
\end{figure*}

We also conducted thematic analysis on the auditing insights authored by the participants. The results are shown in \autoref{tab:insights}. We find that almost all participants paid significant attention to evaluating the \textit{quality} of images, assessing them from pixel-level (e.g., resolution), component level (e.g., human anatomy), to a macro level (e.g., theme and style). We summarize them into three subcategories: \textit{Exquisiteness}, \textit{Authenticity}, and \textit{Style}. Many of them also moved beyond single-image evaluation, focusing on multi-image features like distribution and diversity. 

Another common focus is \textit{Prompt Interpretation}, where participants investigated if the generated images aligned with the prompt, common sense, or their understanding. This reveals the gap between the knowledge of existing T2I models and both the objective world and the specific auditor. 

Furthermore, a few auditors mentioned their reasoning for the LLM's behavior in their insights. With the help of visualizations in Vipera, like stacked bar charts, participants were enabled to compare images and their distributions from different prompts easily. Some of them summarized the LLM's tendency in image creation or even explained it, proposing assumptions about the LLM's training sets. 

Generally, we found the multi-modal guidance employed in Vipera enhanced the insight diversity in the audit reports, as illustrated by an example case shown in \autoref{fig:example}. For instance, P21 initially focused on assessing the image quality, vibe, and race diversity in System A. The criticism of the model's tendency to generate artistic images remained in the report from System B, but he also mentioned the issue within the diversity of blanket colors, citing a chart from the scene graph. In addition, the LLM-driven guidance in System D further inspired him to find that all images were taken from the front, which could be improved for better diversity.  

It's worth noting that although we clarified in the system tutorial that the task was to audit the T2I model specifically, a few participants still misunderstood the goal and commented on other system components, such as the scene graph and the AI labeler. For instance, as highlighted in \autoref{fig:example}, P21 included a sentence commenting on the labeling quality in the report for System A.  Relatedly, while many participants noticed the errors of the AI labeler, only about half of them tried to correct the errors or asked for relabeling with modified expressions, with many participants ending up trusting the wrong results. This pattern extends prior work on aligning user intents with LLM-based judges~\cite{who-validates, pan2024human}, showing that even when users are prompted to evaluate a specific component, failures in automated judges can go unchallenged, so such tools should be treated and deployed with caution.

\subsection{Qualitative Findings from Interviews} \label{result:interview}

Our main insights from the semi-structured interviews are summarized below.

\subsubsection{\textbf{The scene graph encourages systematic auditing while bringing additional cognitive demand and pressure.}} \label{scene-graph-finding}

We received divided opinions on the effect of the scene graph. Participants in favor of this visual guidance suggested that it provided them with a general overview of the images, served as a good source of inspiration, and encouraged systematic thinking and attention to detail, ultimately making them feel more productive and confident. For instance, P1 noted, ``\textit{I was used to focusing on the macro scenario in the image and treating it as a whole, but the scene graph made me realize that images could be decomposed into individual elements.}'' P21 similarly mentioned that the scene graph gave him ``\textit{a systematic understanding of the images' compositions}''. Moreover,  we observed that some participants acquired inspiration from interacting with the scene graph. P3 mentioned that apart from the tree structure and the node contents, the results of open labeling (i.e., without predefined candidate values) could also be inspiring, reminding him of ``corner cases''.  
However, others found the scene graph's complexity to be a barrier, citing a steep learning curve and information overload. This structured approach also had a psychological cost for some. P06 and P09, for instance, felt pressured by the graph, stating it limited the ``free-form authoring'' he preferred and undermined his sense of agency.  Additionally, several participants expressed neutral views. P21 and P08 argued that the usefulness of the scene graph was most prominent when the prompts  or images were abstract (i.e., with unclear scenes or in an artistic style), which were easier to assess using the scene graph compared with direct inspection. P21 further indicated that the scene graph's usefulness was content-dependent, asserting that nodes related to human characters or biases were more beneficial than those focusing on minor details. \looseness=-1 

\subsubsection{\textbf{AI-powered auditing suggestions, though inspiring and effort-saving, needed to be personalized and upskilled.}} \label{ai-suggestion-finding}

While most participants agreed on the inspiration and saved effort from AI, they suggested various opportunities for improvement. Some participants (P02, P09, P11, P12, P22) believed that the performance of AI auditing support might decrease over time, with the LLM's creativity ``converged'' at some point, providing recommendations that were repetitive, not interesting, or did not make sense. P1 and P21 noted that the LLM tended to modify the original prompt too much when suggesting new prompts, causing significant changes to the images and hence hindering comparison. Others expressed their desire for more personalized suggestions that take the system state and interaction history into account. In addition, we noticed that while Vipera allowed users to customize LLM suggestions through selected keywords or topics, few participants viewed this feature as useful. P07 and P20 described the process of manually aligning the LLM as demanding and reactive. By contrast, they preferred a more proactive, LLM-driven approach that anticipates their preference and adapts automatically.

\subsubsection{\textbf{Effective auditing relies on drawing on diverse sources of insight and integrating them to perform a comprehensive assessment.}}\label{blending-guidance}

Based on the interviews and our inspection of the task process, we have observed six major sources of insights that motivate user interactions during the task: direct image inspection, the scene graph, the AI auditing support, comparing results from different prompts, personal experience, and serendipity. Almost all of them were commonly integrated by all participants with no significant differences between general and expert auditors. Contrary to our assumption that expert auditors may prefer advanced auditing techniques, P22, an expert auditor, highly relied on direct inspection. He shared that ``\textit{examining the images is super effective...If I were asked to start by using the second (C) or third system (D) without actually trying the first (A), I would probably just rely on AI guidance without noticing the insights from checking the images, since I am a lazy person.}'' 

\subsubsection{\textbf{The auditing approaches varied among participants, with both breadth-oriented and depth-oriented patterns in prompts and criteria.}} \label{audit-pattern} Participants showed various patterns during their auditing process and expressed distinct rationales in the interview. Most participants followed a pattern where they attempted a diverse range of criteria on only a few prompts. 
For instance, 9 participants used fewer than 2 prompts on average among all three systems, with four of them spending most time exploring distinct criteria.  P3 explained that he began the task with several rough auditing criteria in mind, so he tended to exhaust those criteria on existing images before moving on to new ones. P18 shared that she found the  images from the initial prompt to be of extremely low quality, leading her to spend significant time evaluating them thoroughly without time for exploring other prompts. 
By contrast, some participants followed another approach, iteratively refining prompts based on new insights in a manner similar to prompt engineering. P5 and P6 explained that, rather than assessing images from multiple perspectives, they tried to obtain satisfactory images after finding the initial outputs unsatisfactory and therefore focused on how model behavior changed across successive prompts. Although some participants ultimately obtained images they were happy with, P6 did not and became frustrated, noting down his constraints in the auditing report. We also noticed one participant, P1, who tried to balance between breadth and depth when exploring prompts and criteria. He noted, ``\textit{whenever I felt that I dived too much in a specific direction, I told myself to think about other ways}''. 



\section{Limitations} \label{limitation}
Our study has several limitations. First, we have identified several threats to the validity of our study. For instance, our controlled experiment, conducted in a lab-like setting with a fixed duration, may not fully capture the open-ended and long-term nature of real-world auditing tasks. Meanwhile, our participant pool consisted primarily of student auditors, whose workflows may differ from those of professional red teamers or compliance officers. The generalizability of our findings could be strengthened by future longitudinal deployments with a more diverse range of experts and domain-specific prompts. Furthermore,  we used bookmarks in our analysis as a proxy for insights, a metric that future work could refine. The fixed order of system presentation could also introduce potential learning or fatigue effects, as the difficulty of finding new insights naturally increases over the course of a session. In addition, the LLMs used in the study could make mistakes that misled participants, which could influence user responses and auditing results. Finally, while Vipera can be extended to audit multiple T2I models in parallel, doing so likely requires further work to adapt guidance and reconcile differences across models.

\section{Discussion}

Based on our results, this section discusses opportunities to enhance the support for the design, development, and evaluation of AI auditing tools that facilitate more systematic auditing. In particular, we highlight how visual guidance can scaffold auditors before, during, and after their work by making complex results more interpretable and actionable. We also examine how AI-driven inspirations and guidance can be designed to better adapt to users’ goals, drawing on our empirical findings. Additionally, we provide insights into best practices for integrating various forms of guidance to support human–AI collaboration in auditing. Finally, we discuss the potential to embed tools like Vipera into on-the-ground responsible AI workflows in industry settings and beyond. Together, these opportunities outline design directions for building future auditing tools capable of scaling to the complexity and diversity of generative AI systems, while enhancing and complementing auditors’ capacity of sensemaking AI outputs at scale.

\subsection{Opportunities from involving visual guidance in algorithm auditing} 

Prior works on AI auditing have explored various types of guidance, ranging from community-based to algorithmic and expert~\cite{everydayaudit, deng2025weaudit}. Our findings reveal the numerous opportunities that arise from providing \textit{visual} guidance. Specifically, we demonstrate that users can benefit from visual guidance throughout the auditing process. Before auditing, it can help users organize their thoughts and provide inspiration (see Section \ref{scene-graph-finding}). During auditing, it can serve as an interactive environment for seamless auditing. After auditing, it can organize the results for systematic analysis. Furthermore, we demonstrate the effectiveness of augmenting auditing with a scene graph in improving user confidence and production. 

Beyond the auditing literature, visual guidance has been widely studied and applied in data analysis and sensemaking tools by both the HCI and visualization community \cite{suh2023sensecape, jiang2023graphologue, guo2024prompthis}. Consistent with these works, we find that visual guidance facilitates deep data understanding and explorative navigation, yet it may introduce risks such as increased cognitive load and information overload. Meanwhile, we have identified several unique design implications. First, we observe that the risks associated with visual guidance can be partially mitigated by LLM-driven guidance, as LLMs can summarize visuals and provide actionable suggestions. Second, given the mixed feedback on the scene graph from participants, we believe that visual guidance should be applied contextually, depending on factors such as user preferences, visual literacy, and data/task complexity. Future systems can adapt guidance modality to user profiles and to predicted user intent or task. Third, our study highlights users' diverse strategies for using visual guidance. For instance, given the huge auditing space, participants revealed different behaviors when balancing exploration and exploitation (see Section \ref{audit-pattern}). We envision future design interventions to support various user types, such as a ``focus'' mode for depth-oriented users and a ``discover'' mode for breadth-oriented users. For instance, future interfaces can determine the user type based on how frequently users create nodes in the same branch or cross branches, and take advantage of this information to improve the algorithm for AI guidance generation.

Additionally, given the effectiveness of visual guidance in Vipera, future auditing interfaces could explore a more diverse range of visual designs that better address user needs.
For instance, visualizations could take the form of timelines to capture changes across iterative audits, given similar practices in visualizing iterations of data or code \cite{data-iteration-vis,loops,wang2022diff}.
Auditing interface could also leverage keyword or topic visualizations such as word cloud \cite{heimerl2014word}, cluster maps, or document cards \cite{strobelt2009document} to reveal patterns across diverse user reports. Finally, there are opportunities where existing theories and techniques for visualization recommendation (e.g., \cite{visrec}) or proactive visual analytics support (e.g., \cite{zhao2025proactiveva, weng2025insightlens}) can be integrated, augmenting the auditing process with guided visual data analysis. 

\subsection{Leveraging AI for inspirations and guidance in AI auditing} 

Building on insights from how participants perceived and interacted with AI-generated auditing guidance (see Section \ref{ai-suggestion-finding}), our study extends existing suggestions (e.g., \cite{everydayaudit, deng2025weaudit, enduseraudits}) for designing AI-powered auditing guidance and suggests several advanced lessons. 

First, our finding highlighted that AI-generated auditing guidance should be personalized and built upon the inference of users' goals or preferences. Future design could consider taking advantage of users' interactions, learning from users' feedback, and providing guidance that matches users' focus, as demonstrated in prior literature \cite{heer2019agency, shen2025prompting}. \looseness=-1

Second, a number of participants suggested that AI-generated guidance should incorporate mechanisms to avoid repetitiveness and maintain novelty and relevance over prolonged use. In future interaction design, users should be informed when guidance on auditing criteria (or prompts) is going to converge so that they can try to generate new prompts (or criteria) purposefully. Alternatively, higher sampling temperature or nucleus sampling could be used periodically to surface diverse suggestions. 

Finally, as we found that AI-generated guidance still lacks high-level rationales, effective auditing strategies should be injected into the training process of LLMs beyond relevant knowledge. They can benefit from learning expert practices, such as slightly varying the prompt with controlled variates at each time. This necessitates studies on effective auditing techniques as well as the development of tools capable of eliciting such strategies. 

\subsection{Integrating different types of guidance for effective human-AI collaboration} 

Our findings indicate that combining visual and LLM-driven guidance is more effective than either modality alone for reducing user workload and improving auditing performance (Section \ref{blending-guidance}). These modalities play complementary roles and mitigate each other’s weaknesses: visual guidance conveys concrete, situational information (e.g., UI state, highlighted anomalies, and semantic relationships among auditing criteria) that helps both users and models ground their reasoning, while LLM-driven guidance synthesizes context, generates high-level hypotheses, and proposes next-step strategies.
Specifically, the LLM can reduce the additional cognitive load introduced by visual guidance by summarizing and prioritizing visual information; conversely, visual guidance helps the LLM and users by communicating system state and on-screen evidence in an immediately perceivable form. Together, they streamline iterative sense‑making: visual cues attract attention and reduce search cost, while textual LLM explanations contextualize those cues, translate model inferences into actionable audit steps, and justify recommendations in familiar language.

Practically, this multimodal loop improves transparency and trust by making the AI’s reasoning traceable (users can link suggestions to visual cues), supports progressive disclosure of complexity (e.g., intricate criteria can be hierarchically represented and revealed through the scene graph), and accommodates diverse user preferences and expertise levels. To fully realize these benefits, system designs should tightly couple the modalities, such as enabling cross‑modal interactions like clicking a visual highlight to obtain an LLM justification. Future work could draw from recent systems  \cite{zhao2025proactiveva, self-regulated-learning-vis} to design such interactions.

\subsection{Incorporating intelligent auditing system into real world}

Through designing, developing, and evaluating Vipera, our work provides initial empirical evidence and design implications for creating better auditing tools that can systematically surface and support sensemaking of potentially problematic AI-generated image outputs (see Section \ref{result:mixed-method} and \ref{result:interview}). However, as noted in our Limitations section (see \ref{limitation}), future work is much needed to evaluate Vipera in more ecologically valid environments with real-world auditors in industry settings and beyond, aligning with many calls from prior work on studying responsible AI practices in real-world contexts~\cite{balayn2023fairness, kaur2020interpreting}. We believe future deployments of Vipera could fit into organizational auditing workflows in three key ways. First, to better situate Vipera in current developer workflows in light of the calls from recent RAI tool development in HCI~\cite{deng2022exploring, wang2024farsight}, it could be integrated into Jupyter notebooks to reduce adoption friction. Second,  Vipera's visual analytics capabilities could assist red-teaming pipelines by providing a structured environment for testing adversarial prompts. Furthermore, the insights derived from Vipera could be directly linked to model cards or dataset documentation, potentially serving as a forum for auditors to associate granular findings with specific models or training data.

In addition, because AI auditing and other RAI practices should be collaborative efforts among cross-functional teams~\cite{deng2023investigating, wang2023designing, passi2019problem, holstein2019improving}, future researchers could extend Vipera and similar auditing interfaces to support communication across different roles. For instance, inspired by prior work in HCI and RAI~\cite{yildirim2023investigating, bogucka2024co}, the ``Note view'' feature (see Section \ref{vipera:note}) could be more explicitly designed to help auditors or model evaluators communicate auditing results to business analysts or leadership without technical backgrounds. 

Finally, a large body of HCI and RAI research has shown that practitioners often encounter pushback from leadership when advocating for more responsible technologies~\cite{madaio2020co, rakova2021responsible, kumar2024balancing}. Moreover, despite the growing number of RAI tools, organizational studies of industry practices highlight how the profit-driven and fast-paced nature of industry work frequently discourages meaningful RAI engagement~\cite{widder2024power, ryan2024ai}. While our results demonstrate greater time efficiency for auditors (see Section \ref{result:mixed-method}), future work should explicitly examine how tools like Vipera would be socially situated within organizational contexts, and what roles computational tools for systematic GenAI auditing can—and should—play within broader RAI efforts in industry.

\subsection{Future work}
Additionally, we have identified several promising directions for future research. 

First, participants expressed a desire for more personalized and proactive guidance. Future versions could incorporate a user model that learns an auditor's focus to anticipate their needs and surface more relevant suggestions. 

Second, our observation that users often did not correct errors made by the AI labeler points to a need for improving human-AI collaboration. Future interfaces could integrate confidence scores, label explanations, and more intuitive correction mechanisms to foster greater user engagement and trust. In line with prior work, such as EvalGen by Shanker et al. \cite{who-validates}, future research is needed to develop human-in-the-loop validation mechanisms that ensure the accuracy and truthfulness of LLM-driven guidance. In addition, before deploying Vipera, practitioners could construct benchmarks to evaluate the accuracy of scene-graph generation and better understand the reliability of the LLM outputs. Such measurements would greatly improve auditors’ calibration of how much to rely on LLM-generated guidance. \looseness=-1

Third, Vipera is designed for individual auditors to create their own auditing report. Given that modern auditing activities often involve multiple auditors~\cite{everydayaudit}, developing tools that systematically integrate multiple reports into a final, cohesive report is a promising direction.

Finally, Vipera could be enhanced by expanding the granularity of auditing guidance and cross-modal interactions, such as linking anomalies in charts to the pixel-level areas within the raw image data to create a more fine-grained sensemaking loop. 

\section{Conclusion}
We introduce Vipera, an innovative system designed to enhance the systematic auditing of these models by providing structured, multi-faceted analysis and LLM-powered suggestions. Vipera further leverages multiple visual aids, including scene graphs and stacked bar charts, to facilitate intuitive sensemaking of auditing results and assist auditors in streamlining and organizing their auditing. Our user study confirms Vipera's usability and effectiveness, demonstrating its potential to help auditors navigate the complex auditing landscape while uncovering new insights. The study also demonstrates the benefits of blending both visual and LLM-driven guidance, with participants showing various preferences and behavioral patterns. We believe Vipera will contribute significantly to the advancement of end-user auditing and foster more responsible human-AI collaboration in creative applications.

\begin{acks}
The research was supported by National Science Foundation (NSF) program on Fairness in AI in collaboration with Amazon under Award No. IIS-2040942, as well as an award from Notre Dame–IBM Technology Ethics Lab. We would like to thank all the anonymous reviewers for their constructive comments.
\end{acks}

\bibliographystyle{ACM-Reference-Format}
\bibliography{main}










\end{document}